# Modification of low-temperature silicon dioxide films under the influence of technology factors

B. I. Seleznev

Yaroslav-the-Wise Novgorod State University, Veliky Novgorod, 173003, Russia

*Boris.Seleznev@novsu.ru*

The structure, composition and electrophysical characteristics of low-temperature silicon dioxide films under influence of various technology factors, such as ion implantation, laser irradiation, thermal and photonic annealing, have been studied. Silicon dioxide films have been obtained by monosilane oxidation using plasma chemical method, reactive cathode sputtering, and tetraethoxysilane pyrolysis. In the capacity of substrates, germanium, silicon, gallium arsenide and gallium nitride were used. Structure and composition of the dielectric films were analyzed by methods of infrared transmission spectroscopy and frustrated internal reflectance spectroscopy. Analysis of modification efficiency of low-temperature silicon dioxide films has been made depending on the substrate type, structure and properties of the films, their moisture permeability, dielectric deposition technique, type and dose of implantation ions, temperature and kind of annealing.

**Introduction**

Low-temperature $SiO_2$ films are widely used in technologies for development of microdevices based on silicon, gallium arsenide, nitrides of III group [1, 2]. During creation of microdevices, $SiO_2$ films are affected by influence of different technology factors, for instance, ion beams, moisture, annealing, plasma. Analysis of influence of various technology factors on structure and parameters of silicon dioxide films is therefore of practical interest. Quest for ways of parameters control of low-temperature $SiO_2$ films is an actual problem. In **[3]**, control of stoichiometry and structure of low-temperature plasma-chemical silicon dioxide was carried out by changing the ratio of the reaction components. Formation of modified areas with changed optical properties in bulk of porous glass sheets was considered in **[4]**. A possibility for controlling of luminous properties of thermally grown $SiO_2$ layers by implantation of boron, phosphorus and carbon ions was demonstrated in [5].

Problems of parameter management of low-temperature $SiO_2$ films can be solved using a combined effect of "soft" laser radiation, ion beams of various kinds, plasma or photon treatments.

In this paper, we analyze the efficiency of modification of low-temperature silicon dioxide films, depending on the type of substrate, the structure and properties of the films, their moisture permeability, the methods of forming a dielectric, the type and dose of the ions being introduced, the temperature and the type of annealing.

**Initial samples and experimental procedure**

Germanium, silicon, gallium arsenide and gallium nitride were used as semiconductor substrates. Silicon dioxides on semiconductor substrates were fabricated by various chemical and ion-plasma techniques: $SiO_{2pch}$ – by plasma-chemical assisted vapor deposition, $SiO_{2r}$ by reactive cathode sputtering, $SiO_{2pyr}$ by pyrolysis of tetraethoxysilane, $SiO_{2sil}$ by oxidation of monosilane in oxygen atmosphere. $SiO_{2pch}$ films were fabricated by plasma-chemical deposition in inductively coupled plasma using Sentech SI 500 D technique. Preliminary pumping of the vacuum chamber was at least $5·10^{-7}$ mm Hg. The films were deposited in vacuum at least $5·10^{-2}$ mm Hg at temperature 200 $^0$C.

$SiO_{2sil}$ films were prepared by monosilane oxidation using a setup for pyrolytic deposition of low-temperature dielectric films. Monosilane and oxygen were used as precursors during the deposition of $SiO_{2sil}$ films. Argon was used as gas hearer. The process was conducted at temperature 250 $^0$C. $SiO_{2pyr}$ films were deposed by the way of pyrolysis of tetraethoxysilane using "Isotron" technique by open-tube method at temperature 680 $^0$C. For comparison, silicon nitride films obtained by plasma chemical method and magnetron sputtering of $Si_3N_{4pch}$, $Si_3N_{4ms}$ were used.

Setup As-One for high-temperature photon annealing was used for impurity activation during ion implantation. The setup reaches temperature 1500 $^0$C with the rate of temperature increase 200 $^0$C/sec. Annealing was conducted in nitrogen atmosphere.

$SiO_2$ films were implanted by $B^+$, $BF_2^+$, $Ar^+$, $As^+$, $P^+$ ions with doses $10^{12} – 10^{17}$ cm$^{-2}$. Radiation treatment with wavelength 1.06 μm in free oscillation mode was used in the experiments. Selection of this wavelength of laser radiation was caused by that the low-temperature dielectric films often contain large number of water molecules and OH groups. This leads to weakening of the films under laser treatment in short wave range in comparison to radiation treatment with wavelength 1.06 μm [6]. Laser annealing was conducted at pulse energy $E_p = 0.2–13$ J/cm$^2$, $\tau_p = 0,8$ ms.

Structural changes in the dielectric films were estimated by Fourier infrared spectrometry and multiple frustrated total internal reflection (MFTIR) [7]. FSM 1202 infrared Fourier spectrometer was used for this task. Samples of n-germanium of trapezoidal shape with specific resistance 20 Ω·cm possessing the transparency in required range of wavelength were used for studying of structure and composition of $SiO_2$ films (Fig.1). At single pass of light beam through the MFTIR element with length *l* equal to one a total number of reflections is N = (l/t)·cotan θ. At given parameters *l* = 50 mm, *t* = 0.8 mm, number of reflections is equal to N = 60.

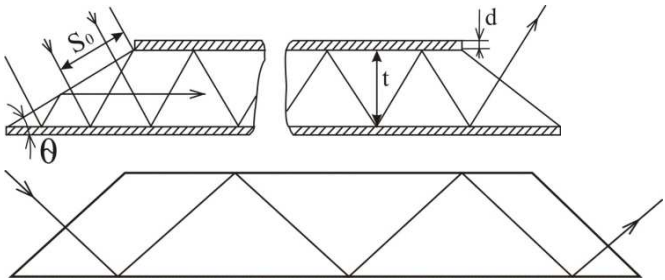

Fig. 1. MFTIR element of single pass

Fig.2. Sample of MFTIR comparing

Germanium sample with small number of reflections (5) was used as comparing sample. It enables to exclude the influence of structural features of comparing sample surface on shape of MFTIR spectra, see Fig.2. To analyze the glasses of complex composition in the range ~ 1325-1380 cm$^{-1}$, high-resistance silicon with specific resistivity 20 KΩ·cm was used as MFTIR sample. Electrophysical properties of systems "dielectric-semiconductor" were estimated from measurements of electrical strength of the films using a method of nondestructive examination for noise characteristics [8]. Thickness of dielectric films was measured by ellipsometric characteristics. Scanning electron and atomic force microscopy were used for analysis of influence of high-temperature annealing on the structure of the "dielectric-semiconductor" samples.

**Structure and composition of initial SiO$_2$ films**

Fig.3 shows IR transmission spectra for SiO$_2$ films fabricated by various techniques. SiO$_{2therm}$ films, prepared using thermal oxidation of silicon in chlorine-containing media at *T*= 1050°C, were used as etalon films. Thickness of SiO$_2$ layers for all three experiments was about 0.4 μm.

Analysis of main transmission band in the range 1060–1090 cm$^{-1}$ connected with valence vibrations of Si–O bonds shows that the films fabricated by plasma-chemical method have the best quality in comparison with the films prepared by oxidation of monosilane with oxygen (namely, half-width of the band, position of transparency maximum, band intensity).

As it follows from Fig.3, spectra of IR transparency for SiO$_2$ films have a transmission band near 883 cm$^{-1}$ due to lack of oxygen (Si$_2$O$_3$).

Photon annealing was used to eliminate the oxygen deficiency. After the photon annealing at *T* = 500 °C, *t* = 30 s the absorption band near 883 cm$^{-1}$, caused by oxygen deficiency, disappears (Si$_2$O$_3$). At that, minimum of the main transparency band shifts to high frequency side from

1066 cm$^{-1}$ to 1073 cm$^{-1}$ due to the transformation Si$_2$O$_3$ into SiO$_2$ and the band intensity increases. The increase of the annealing time from 30 s up to 10 min does not significantly change the intensity of the transparency band near 883 cm$^{-1}$.

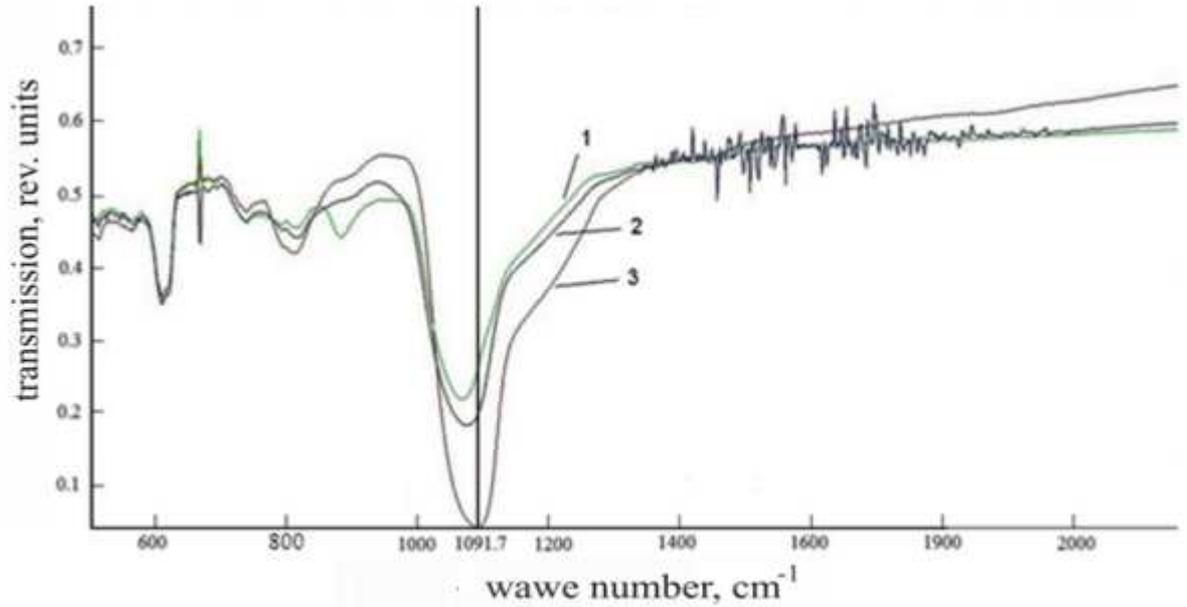

Fig. 3. Transmission spectra of dielectric SiO$_2$ films on silicon substrates fabricated by different techniques:   1– SiO$_{2sil}$; 2 – SiO$_{2pch}$; 3 – SiO$_{2therm}$

Fig.4 shows the MFTIR spectra for Ge–SiO$_{2sil}$ samples. Just after the sputtering of SiO$_{2sil}$ films an intense reflection band appears in the MFTIR spectra which is typical for adsorbed molecules and hydroxyls in oxide, see Fig.4a [9].

The presence of water in the SiO$_{2sil}$ film was confirmed by presence of a reflection band in the range 1640 cm$^{-1}$ of the spectrum in Fig.4b, which is due to deformation vibrations of molecularly absorbed water. A shoulder near 940 cm$^{-1}$, appearing in the MFTIR spectra after holding of Ge-SiO$_{2sil}$ samples in air (Fig. 4b) caused by deformation vibrations of Si–OH groups indicates that water had reacted with SiO$_2$ and formed Si–OH bonds. Reflection band at 870 cm$^{-1}$, found also in the IR transmission spectra, can be related to formation of intermediate oxide Si$_2$O$_3$ during the film application. High intensity reflection band in the OH region indicates a high degree of the films porosity. Apparently, a degree of deformation of Si – OH bond, together with porosity, is high in the SiO$_{2sil}$ films [10]. When the water reacts with the oxide to form Si–OH groups, the deformation of the bond weakens and the absorption in the main band increases, see Fig. 4b, curve 2.

For SiO$_{2pyr}$ films, an intensive reflection band, typical for hydroxyls in oxide and adsorbed water molecules, was observed in the range 2700-3700 cm$^{-1}$. SiO$_{2r}$ and SiO$_{2pch}$ films are the hydrophobic ones. They characterized by only reflection band with a minimum near 3630 cm$^{-1}$ caused by oscillations of Si–OH groups.

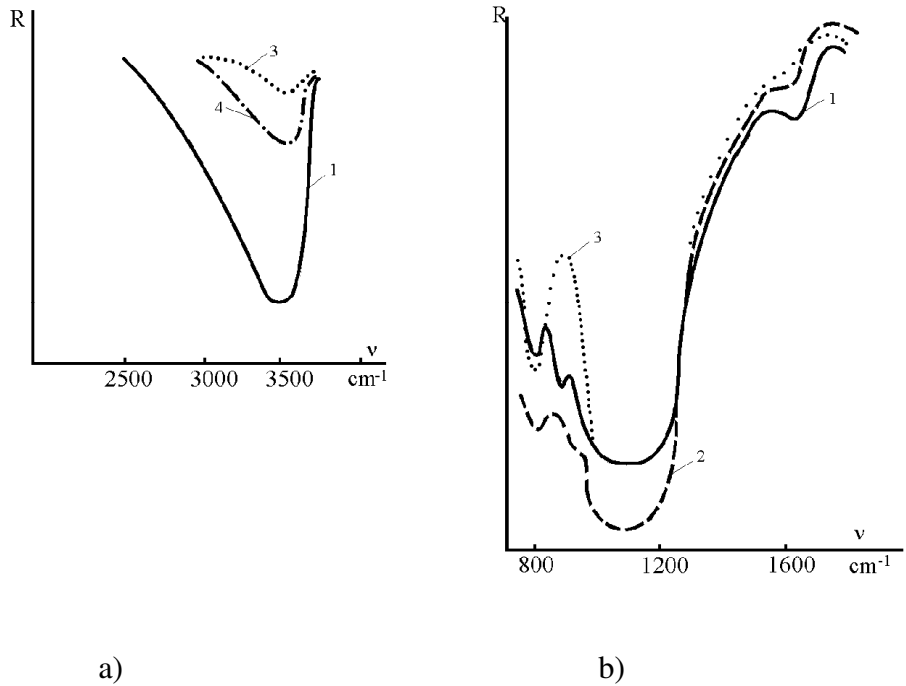

a)            b)

Fig.4. MFTIR spectra for Ge–SiO$_{2sil}$ sample: 1 – just after application of the SiO$_{2sil}$ film; 2– after the holding in air during 17 days; 3 – after annealing in vacuum at 500°C; 4 – after annealing in vacuum at 500 °C and holding in air during 6 days.

**Implantation and laser annealing of SiO$_2$ films**

Irradiation of non-implanted samples Ge–SiO$_2$, Si – SiO$_2$ by neodymium laser pulses at radiation energy density 1–10 J/cm$^2$ did not lead to observable structural changes of SiO$_2$ films. Structure improvement under laser annealing was observed only for SiO$_{2sil}$ films characterized by high porosity and significant deformation of Si–O bonds. Influence of annealing under laser treatment was not observed for hydrophilic SiO$_{2pyr}$ films with low degree of bond deformation. Therefore, radiation treatment of SiO$_{2sil}$ films on germanium and silicon with millisecond pulses of neodymium laser reduces the deformation of Si – O bonds. Under energy densities higher than 10 J/cm$^2$, observable structural imperfections and microcracks arise in SiO$_2$ films.

Radiation treatment with millisecond pulses of neodymium laser of non-implanted $SiO_{2pch}$ and $SiO_{2sil}$ films on gallium arsenide did not lead to observable structural changes in the films at energy density of a pulse below 4 J/cm². Higher energy densities of a pulse result in breaking of Si–O–Si bridge bonds.

In case of implantation of $Ar^+$, $B^+$, $BF^+$, $P^+$, $As^+$ ions in $SiO_2$ films, annealing of irradiation-induced structural defects in these films is efficient for implantation of $P^+$, $As^+$ ions for silicon or germanium substrates. Implantation of ions in $SiO_2$ films leads to breaking of Si–O bonds, change of the angle between Si – O – Si bonds, shift of atoms and formation of new molecular complexes. Structural changes in $SiO_2$ films after ion implantation were displayed in IR transmission spectra and MFTIR spectra (Fig. 5, curves 2, 5; Fig. 6, 7). Irradiation of Si–$SiO_2$ ($P^+$) and Si – $SiO_2$ ($As^+$) samples by millisecond pulses of neodymium laser leads to removal of radiation structural damages fixed by IR transmission spectra and "recovery" of the film structure to the initial form (before implantation), see Fig.5, curves 3, 6.

Probably, effect of laser annealing of radiation-induced structural disorders in $SiO_2$ films implanted by $P^+$, $As^+$ ions can be explained by absorption of laser radiation on microinhomogeneities [11], perhaps, of layered configuration type, which are formed in the films during ion implantation. Submicron particles and inhomogeneities less than 0.1 μm in size absorb weakly and do not influence noticeably the volume of dielectric film where they are contained. However, they can initiate an effective absorption of laser radiation energy in a large volume [12].

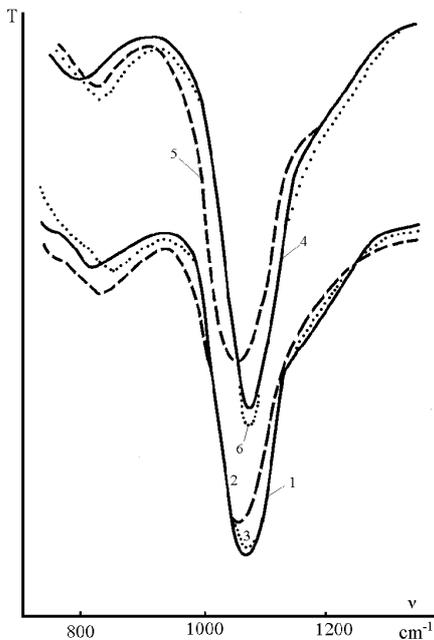

Fig. 5. IR transmission spectra for $SiO_2$ films on silicone implanted by $P^+$ ions with energy 100 keV and dose $1 \cdot 10^{15}$ cm$^{-2}$ (curves 2, 5) and irradiated by millisecond pulses of neodymium laser (curves 3, 6). (1–3) – $SiO_{2sil}$; (4–6) – $SiO_{2pch}$; 3 – $E_p$ = 5 J /cm²; 6 – $E_p$ = 2 J/cm²; 1, 4 – the initial spectra.

Formation of microinhomogeneities in $SiO_2$ films at implantation of phosphorus ions can be related to appearance of P–Si bonds arising at substitution of oxygen atoms by phosphorous atoms between Si – $O_4$ tetrahedrons. In this case, number of P –

Si bonds is well over than the number of P–O bonds formed at substitution of silicon by phosphorus atoms in Si – O$_4$ tetrahedrons.

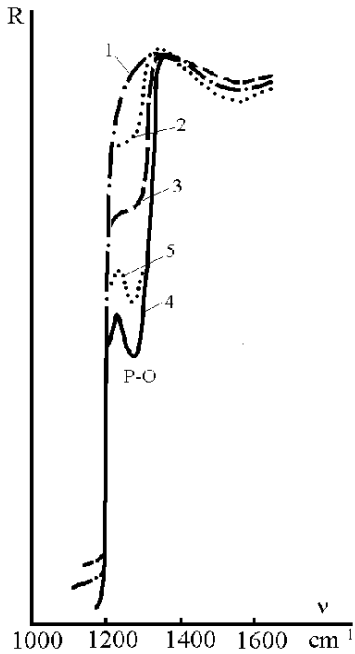

Fig. 6. MFTIR spectra for Si–SiO$_{2pch}$ system
1 − before implantation;
2 – implantation by P$^+$, D = 5 · 10$^{16}$ cm$^{-2}$, E = 50 keV;
3 – implantation by P$^+$ and annealing in oxygen, T = 500$^0$C;
4 – implantation by P$^+$ and annealing in oxygen, T=700$^0$C;
5 – implantation by P$^+$ and laser annealing;
E$_p$=4 J/cm$^2$, τ$_p$=0,8 ms, λ =1.06 μm, film thickness was 0.23 μm.

**Ionic synthesis**

A process of forming of phosphorous-silicate glass on silicon surface was studied by MFTIR method. According to MFTIR IR spectrometry data, small amount of P–O bonds formed just after implantation of P$^+$ ions (Fig. 6, curve 2). P–O bonds arise under annealing of SiO$_2$ films (P$^+$): thermal annealing in oxygen atmosphere or under irradiation by ms pulses of neodymium laser (Fig. 6, curves 3, 4, 5). Annealing leads to remarkable change of intensity of reflection band near ~ 1325 cm$^{-1}$ caused by formation of P –O bonds.

At implantation of BF$_2^+$ and B$^+$ ions in SiO$_2$ films, B – O bonds arise just after the ion implantation and effect of annealing of radiation-induced structural failures was displayed weakly. After ion implantation, reflection band near ~ 1380 cm$^{-1}$ was observed in MFTIR spectra caused by formation of boron oxide (Fig. 7). Note that heating of germanium or silicon substrate is immaterial at laser irradiation with wavelength of 1.06 μm of SiO$_2$ films implanted by phosphorus or arsenic since absorption coefficients, at near the same effectiveness of "regeneration" of the structure of implanted films, differ considerably: α (Si) ≈ 20 cm$^{-1}$, α (Ge) ≈ 2·10$^4$ cm$^{-1}$.

Structure "regeneration" of SiO$_2$ films implanted by phosphorus ions was observed for implantation doses less than 10$^{16}$ cm$^{-2}$. For this case, one pulse makes the sufficient influence.

In case of laser treatment of implanted films $SiO_2$ ($P^+$) and $SiO_2$ ($As^+$) on arsenide gallium, annealing effect was not observed, probably, due to considerable internal mechanical stresses in the system.

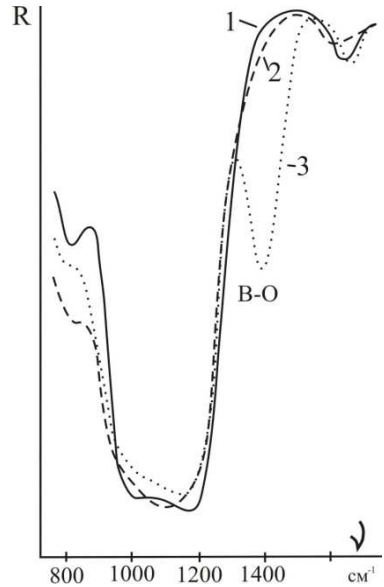

Fig. 7. MFTIR spectra for Ge–$SiO_{2pyr}$ structures.
1 – initial $SiO_{2pyr}$ film; 2 – after $BF_2^+$ implantation, E = 50 keV, D = 1 · $10^{15}$ cm$^{-2}$ ; 3 – after $BF_2^+$ implantation, E = 50 keV, D = 1 · $10^{16}$ cm$^{-2}$ ;

Film thickness was 0.18 μm, number of reflection was 40.

**Electrophysical characteristics of systems "$SiO_2$-semiconductor" irradiated by laser**

In present paper, an analysis of electrophysical characteristics of systems "silicon-implanted $SiO_{2sil}$ film", treated with laser irradiation, was conducted. Doses of $P^+$ ions implanted in $SiO_{2sil}$ films were in the range 1 ·$10^{15}$ ÷ 5 ·$10^{16}$ cm$^{-2}$, energy dose was 50 ÷100 keV. It was possible to decrease the value of fixed charge and hysteresis at pulse energy density of laser radiation of about ~ 4.5 J/cm$^2$. At that, the electrical strength $E_{es}$ of $SiO_{2sil}$ films increases by a factor of 3. At large implantation doses (~ 5·$10^{16}$ cm$^{-2}$) the implanted layer has a "shielding" effect on the characteristics of Si – $SiO_{2sil}$ interface under laser irradiation, and they do not change up to pulse energy densities of laser radiation of the order of ~ 6 J/cm$^2$. Decrease of $E_{es}$ of the films at further $E_p$ increase related to partial crystallization in bulk of the films.

It should be noted that under decrease of implantation dose a maximum arises in dependence $E_{es}$ on energy of laser irradiation both for Si – $SiO_{2sil}$ system and for other types of $SiO_2$ films on silicon or other systems, for instance, SiC – $SiO_2$.

Laser irradiation of GaAs – SiO$_2$ systems or, for comparison, GaAs – Si$_3$N$_4$ system in wide range of energy density did not lead to any improvement of the electrophysical properties of the films. For creation of a "controlled" under laser irradiation GaAs system, it is necessary to form an intermediate layer of intrinsic oxide GaAs. Intrinsic oxide films on gallium arsenide were grown by thermal (TOF), anode (AOF), plasma (POF) and photon (PHOF) oxidation, and on germanium substrate by thermal oxidation.

Pulse laser annealing of systems GaAs – TOF – SiO$_{2r}$ and GaAs – TOF – Si$_3$N$_{4ms}$ (magnetron sputtering) in the wide range of pulse energy density, at variation of thickness of ceramic oxide films, did not lead to an improvement of electrophysical properties of dielectric films as distinct from, for instance, GaAs – TOF – SiO$_{2sil}$ system. Apparently, at ion beam methods of application of dielectric films, radiation-induced damage of surface of gallium arsenide and sputtered films by high energy particles and pollution of the films by atoms or molecules of working gas reduce considerably the effectiveness of laser annealing of systems "GaAs – dielectric".

Wide possibilities for management of electrical strength of dielectric films on gallium arsenide are demonstrated in Fig.8. Use of photon oxide films permits to control E$_{es}$ both for silicon dioxide films (SiO$_{2sil}$) and silicon nitride films (Si$_3$N$_{4pch}$). In case of absence of intermediate intrinsic oxide, laser irradiation has not influence the film characteristics. It should be noted that both thermal and photon annealing in a wide temperature range up to ~800 $^0$C did not lead to a noticeable increase in the electrical strength of the films. Consequently, the possibility of considerable increase of E$_{es}$ for dielectric films on GaAs is a special feature of pulse laser annealing.

The effects of control of the electrophysical characteristics of dielectric-semiconductor systems under neodymium laser irradiation with the wavelength of 1.06 μm in the free-running regime are apparently due to the absorption of laser radiation on microinclusions in the volume of dielectric films and in the dielectric-semiconductor transition layer. Oxygen or free silicon can be the absorbing microinclusions in silicon dioxide and in silicon nitride [13].

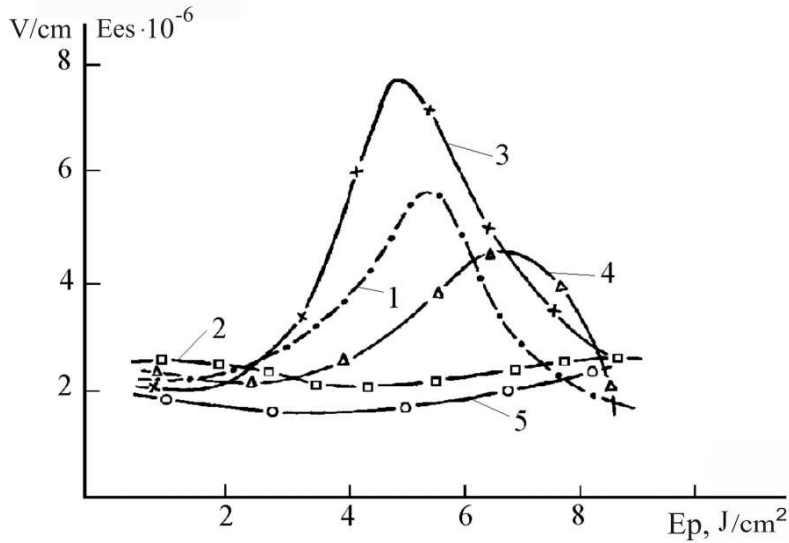

Fig. 8. Dependencies of electrical strength of dielectric films on gallium arsenide on pulse energy density of neodymium laser.

1 – POF– $SiO_{2sil}$ ;  2 – $SiO_{2sil}$ ;  3 – POF–$Si_3N_{4pch}$;  4 – TOF–$Si_3N_{4pch}$ ;  5 – $Si_3N_{4pch}$ ;

Films thickness were: $SiO_{2sil}$ – 0.2 μm, $Si_3N_{4pch}$ – 0.18 μm, POF – 0.02 μm, TOF – 0.03 μm.

The ability to control the electrophysical characteristics of "Ge – dielectric" system can be realized in case of formation of transition layer from germanium dioxide in the system "Ge-$SiO_2$". Laser irradiation of systems Ge – $GeO_2$ – $SiO_2$ leads to improvement of interface properties and stabilization of charge state of the system "germanium-dielectric".

**Protective $SiO_2$ coatings during annealing of implanted GaAs and GaN layers**

When forming ion-doped layer (IDL) on GaAs, an important task is the preparation of layers with high degree of activation of implanted impurity, which is determined by kind of the impurity, conditions of implantation, post-implantation annealing, and quality of initial semi-isolating gallium arsenide. Silicon was taken as dopant impurity for producing of **n**- and **n⁺-n**-layers on GaAs [2].

One of the problems of ion implantation in GaAs consists in dissociation of sample surface under usual temperatures of annealing in the interval 700–900 $^0$C. To prevent this phenomena, in this paper we used $SiO_2$ films as protective coating at IDL annealing on GaAs.

As distinct from the films on silicon, the band at ≈450 см⁻¹ is absent in IR spectrum of $SiO_{2sil}$ films on GaAs. This band is due to deformation doubly-degenerated oscillations of Si–O bonds that can be related to substrate effect. Annealing of GaAs IDL at presence of protective coating $SiO_{2sil}$ leads to significant redistribution of Cr in a subsurface GaAs layer. According to secondary ion mass spectrometry (SIMS) data, chromium concentration after annealing increases sharply near the $SiO_{2sil}$ boundary. One of the most important reasons of such redistribution is presence of mechanical stresses at GaAs–$SiO_{2sil}$ interface. Besides, of the stressed GaAs–$SiO_{2sil}$ boundary, gettering by defects in the implanted layer effects the character of the chromium redistribution.

Besides the chromium redistribution effect on annealing of GaAs-$SiO_{2sil}$ system, another feature of the annealing, according to SIMS data, is diffusion of gallium in $SiO_{2sil}$, see Fig.9.

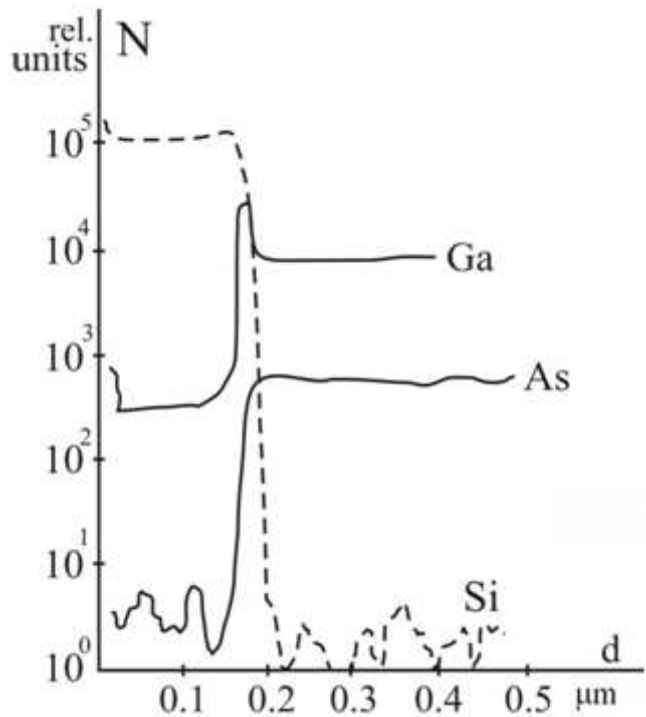

Fig.9. Gallium distribution in a system "silicon dioxide – gallium arsenide" after ion implantation and activating annealing, measured by SIMS technique. Thickness of the $SiO_{2sil}$ film is 0.2 μm.

Increase of surface concentration $n_s$ with growth of annealing temperature from 800 up to 900 $^0$C, in case of presence of protective $SiO_{2sil}$ coating, for implantation doses higher than $2 \cdot 10^{14}$ cm$^{-2}$, can be related to the effect of backward diffusion of gallium and acceleration of diffusion of silicon atoms.

In case of small doses of implanted ions, linear dependence of $n_s$ on dose was observed in interval from $2,5 \cdot 10^{12}$ up to $10^{13}$ cm$^{-2}$ at presence of protecting $SiO_{2sil}$ films. For implantation dose $\approx 10^{13}$ cm$^{-2}$, degree of activation is maximal and reaches $\approx 70 - 80$ %.

The electrical and optical characteristics of gallium nitride layers implanted by silicon were studied in detail in [14]. $^{28}Si^+$ ions were implanted by doses $D= 1 \cdot 10^{13} - 5 \cdot 10^{15}$ cm$^{-2}$ with ions energies $E$=200 keV. Implanted samples were annealed at temperatures $1050-1350$ $^0$C in nitrogen atmosphere using **AlN** coatings**.** Total (100%) activation degree was achieved at annealing temperature **1350 ºC.**

In present paper, creation of sub-contact **$n^+$**-layers on GaN was conducted by implantation of $^{28}$Si$^+$ ions in gallium nitride layers. Next, the structures were subjected to high-temperature photon annealing in nitrogen atmosphere using protecting SiO$_{2pch}$ coatings.

Selection of low temperature SiO$_{2pch}$ films as protecting coatings at annealing of ion-doped GaN layers was determined by high reproducibility of their fabrication, wide application at production of various microelectronic devices, including diode structures based on gallium nitride [1]. Selection of SiO$_2$ as protecting coatings under annealing was determined, according to atomic force microscopy data, its higher thermal resistance in comparison with silicon nitride.

Studied GaN samples were grown by MOCVD method on sapphire substrate (http://ru.wikipedia.org/wiki/MOCVD) of two inches in diameter. Thickness of high resistive active GaN layer was 2.5 μm, and buffer layer – 1.5 μm. SiO$_{2pch}$ films used in the present study were formed by the method of plasma-chemical deposition in an inductively coupled plasma at Sentech SI 500 D unit.

Implantation of Si$^+$ ions in GaN was carried out in various modes. Implantation doses were in the range of $10^{14} - 10^{15}$ cm$^{-2}$, energies of implanted ions were within 50–100 keV. A setup for high temperature photon annealing As-One was used for activation of impurity. Annealing was conducted during 1 min in temperature interval 1100 – 1350 $^0$C in nitrogen atmosphere. Unit for measuring of Hall effect HMS-5000/0/55 T was used for determining of electrophysical parameters of ion-doped GaN layers.

Quality of surface of the structures "gallium nitride- dielectric" subjected for photon annealing was estimated using atomic force microscopy. According to AFM data, the samples annealed at temperature 1200 $^o$C show a small number of dome-shaped defects in the form of brows with height up to 30 nm and with diameter from 20 to 50 nm. Such defects can be associated with accumulations of gallium atoms after the evaporation of nitrogen atoms during annealing process.

In view of presence of defects in the initial gallium nitride substrates, experimental data for implantation doses higher than $10^{15}$cm$^{-2}$ are of practical interest. In case of annealing with SiO$_2$ protective coating, high degree of impurity activation (46-80%) at annealing temperature 1200 $^0$C and degree ~100 % at annealing temperature 1250 $^0$C was observed. Annealing temperatures higher than 1300 $^0$C lead to fracture of dielectric coatings. At the same time, according to atomic force microscopy, clusters of gallium atoms are observed on the surface.

The surface concentration is most stable over a wide temperature range for samples subjected to annealing at a temperature of 1250 °C and characterized by the highest degree of activation of the impurity.

## Conclusions

According to data of multiple broken total internal reflection for $SiO_2$ films, fabricated by both oxidation of monosilane by oxygen and pyrolysis of tetraethoxysilane, an intensive reflection band was observed in the range 2700-3700 $cm^{-1}$ which is typical for hydroxyls in oxide and absorbed water molecules. $SiO_2$ films fabricated by reactive cathode sputtering and plasmochemical method, are more hydrophobic: for them it is only a strip of reflection with a minimum at ~3630 $cm^{-1}$ caused by oscillations of Si–OH groups. In $SiO_2$ films produced by oxidation of monosilane by oxygen, degree of deformation of Si – O bond is high together with degree of porosity. A transmission band in the range 883 $cm^{-1}$ caused by oxygen deficiency ($Si_2O_3$) was observed in transmission IR spectra of $SiO_2$ films. Photon annealing (fast thermal annealing) was used for removal of oxygen deficiency.

Irradiation of $SiO_2$ films by millisecond pulses of neodymium laser permits to control the structure and parameters of the films: to restore the film structure damaged by ion implantation, to form the stabilizing phosphorous silicate glass at the surface of microelectronic structures, to control the electrical strength of the films. To create a system "semiconductor – dielectric" "controlled" under laser irradiation, it is necessary to form an intermediate layer from intrinsic oxide. Annealing effect at irradiation of structures "semiconductor – dielectric" by millisecond pulses of neodymium laser was related to absorption of laser radiation by microinhomogeneities in the film bulk.

Analysis of conditions for formation of ion-doped layers on GaAs and GaN with protecting $SiO_2$ coatings at implantation of silicon ions was conducted. Doses and energies of implanted ions, temperatures of thermal and photonic annealing were determined. Modes for creating of ion-doped layers with high degree of impurity activation were established.

*The work was carried out with the financial support of the Ministry of Education and Science of the Russian Federation within the framework of the project part of the state order (project number is №3.3572.2017).*

# REFERENCES

1. B. I. Seleznev, G. Ya. Moskalev, D. G. Fedorov. On the photon annealing of silicon-implanted gallium-nitride layers. Semiconductors. 2016. Vol. 50. No. 6. Pp.832–838. DOI: 10.1134/S1063782616060221.

2. B.I. Seleznev. Nondestructive techniques for investigating semi-insulating gallium arsenide as initial material for direct implantation //Proceedings of SPIE - The International Society for Optical Engineering 3rd International Workshop on Nondestructive Testing and Computer Simulations in Science and Engineering. Sponsors: Russian Foundation for Fundamental Research, SPIE. St. Petersburg, Russia, 2000, Pp. 308-3015.

3. A.A. Yasunas, D.A. Kotov, O.M. Komar, V.Ya. Shiripov. Optical properties and structure of the low-temperature plasma-chemical silicon dioxide films. 10 International conference «Interaction of Radiation with Solids», September 24-27, 2013, Minsk, Belarus, Pp. 323-325. http://elib.bsu.by/handle/123456789/48466.

4. G. K. Kostyuk, M. M. Sergeyev, E. B. Yakovlev. Formation of modified areas of porous glass saturated with glycerin under the action of laser radiation. Glass Physics and Chemistry. 2013. Vol. 39. No. 5. Pp. 480–489. DOI: 10.1134/S1087659613050118.

5. T.I. Tetelbaum, A.N. Mikhailov, A.I. Belov, A.I. Kovalev, D.L. Vainstein, T.G. Finstad, Y. Golan. Modification of optical properties and phase composition of Si-implanted $SiO_2$ films ion-doped with phosphorous, boron, nitrogen and carbon. Bulletin of the Nizhny Novgorod State University. 2008. No 3. Pp. 40-46. (in Russian).

6. A. M. Miller; L. V. Soustov. Absorption in and laser damage to KDP and DKDP crystals. Sov. J. Quantum Electronics. 1989. Vol.19. No.1. Pp.39–45.
 http://dx.doi.org/10.1070/QE1989v019n01ABEH007693.

7. A. G. Milekhin, C. Himcinschi, M. Friedrich, K. Hiller, M. Wiemer, T. Gessner, S. Schulze, D. R. T. Zahn. Infrared spectroscopy of bonded silicon wafers. Semiconductors. 2006. Vol.40. No.11. Pp.1304–1313. DOI: 10.1134/S1063782606110108.

8. B. I. Seleznev. Degradation processes and instability in microwave field-effect transistors ion-doped layers of gallium arsenide. Vestnik of Novgorod State University. 1999. No. 13. Pp. 55– 60. (in Russian).

9. D. Feijoo, Y. J. Chabal, S. B. Christman. Infrared spectroscopy of bonded silicon wafers. Appl. Phys. Lett. 1994. Vol.65. No. 20. Pp.2548–2550. http://dx.doi.org/10.1063/1.112631

10. W.A. Pliskin. Comparison of properties of dielectric films deposited by various methods. J. Vac. Sci Technol. 1977. Vol. 14. No 5. P. 1064-1081.

11. I.V. Aleshin, S.I. Anisimov, A.M. Bonch-Bruevich, Ya.A. Imas. Optical breakdown of transparent media containing microinhomogeneities. Sov. Phys. JETP. 1976. Vol. 43, No. 4. Pp.631–636.

12. S.I. Anisimov, B.I. Makshantsev. Role of absorbing inclusions in the optical breakdown of transparent media. Sov. Phys. Solid State, 15(4), 735-743 (1973).

13. A V. Butenin, B. Ya. Kogan. Nature of cumulative laser damage to optical materials. Sov. J. Quantum Electron. 1990. Vol.20. No.2. Pp.187–189. https://doi.org/10.1070/QE1990v020n02ABEH005580.

14. J.A. Fellows, Y. K. Yeo, M.-Y. Ryu, R. L. Hengehold. Electrical and optical activation studies of Si-implanted GaN. J. Electronics Materials. 2005. Vol.34. No.8. Pp.1157–1164.